Review article

Domna G. Kotsifaki* and Síle Nic Chormaic

# Plasmonic optical tweezers based on nanostructures: fundamentals, advances and prospects

**Abstract:** The ability of metallic nanostructures to confine light at the sub-wavelength scale enables new perspectives and opportunities in the field of nanotechnology. Making use of this unique advantage, nano-optical trapping techniques have been developed to tackle new challenges in a wide range of areas from biology to quantum optics. In this work, starting from basic theories, we present a review of research progress in near-field optical manipulation techniques based on metallic nanostructures, with an emphasis on some of the most promising advances in molecular technology, such as the precise control of single biomolecules. We also provide an overview of possible future research directions of nanomanipulation techniques.

## 1 Introduction

Optical trapping – the ability to manipulate small particles using light – was first recognized by the awarding of a Nobel Prize in Physics in 1997 to Steven Chu, Claude Cohen-Tannoudji and William D. Phillips for atom trapping and laser cooling techniques [1, 2]. However, optical manipulation has a broad range of applications aside from physics [3], such as biophysics [4], nanoscience [5] and the life science [6, 7]. In 2018, after more than 30 years of steady development as a powerful tool in biological and medical science [8], Arthur Ashkin was awarded half of the Nobel Prize in Physics for his seminal work on "optical tweezers and their application to biological systems." In between the awarding of these two prizes in this topic, the research community witnessed significant progress in advanced optical micro-/nanomanipulation techniques based on optical tweezers (OT). The first application of optical trapping in biology was demonstrated in 1987, when an individual tobacco mosaic virus [9], an *Escherichia coli* bacterium [9], and a live single cell [10] were trapped. Since then, OT have attracted more and more attention due to their unique non-invasive characteristics. A schematic illustration of the historical development of OT is shown in Figure 1. The ability to apply pico-newton (pN) forces to micron-sized dielectric particles, while simultaneously measuring displacement with nanometer-level precision, is now a routine process. However, conventional OT experiments have major drawbacks when used with nanoscale particles, due to the diffraction limit of the focused trapping laser spot size [20, 21], thereby preventing accurate trap confinement. In this regime, the magnitude of the gradient force [11] drops dramatically as the particle size is reduced. Additionally, Ashkin demonstrated that trapping laser powers as high as 1.5 W could trap nanosized particles with diameters of 9–14 nm [11]. However, at such high laser powers, the trapped specimens typically undergo rapid optical damage. These limitations make the trapping and manipulation of very small particles particularly challenging.

Alternative, OT configurations have, of course, been explored and they have shown a great improvement in the trapping quality by, for example, taking advantage of evanescent fields [21, 22]. For example, photonic crystal OT-integrated with microfluidic systems to create a lab-on-chip-platform-have been used to demonstrate sorting and/or manipulation of particles [23–29]. Photonic crystals are near-field nanostructures with a periodic pattern in dielectric properties. These structures can be integrated with microfluidic systems to create a lab-on-a-chip platform

*Corresponding author: Domna G. Kotsifaki,** Light-Matter Interactions for Quantum Technologies Unit, Okinawa Institute of Science and Technology Graduate University, Onna, Okinawa 904-0495, Japan, e-mail: domna.kotsifaki@oist.jp.
https://orcid.org/0000-0002-2023-8345
**Síle Nic Chormaic:** Light-Matter Interactions for Quantum Technologies Unit, Okinawa Institute of Science and Technology Graduate University, Onna, Okinawa 904-0495, Japan.
https://orcid.org/0000-0003-4276-2014

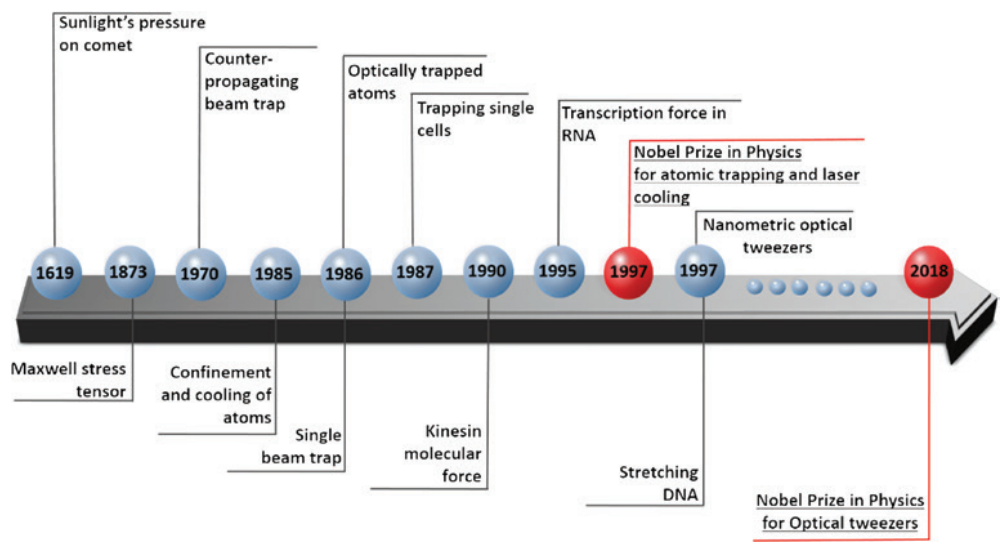

**Figure 1:** The optical tweezers roadmap [1, 2, 9–19].

and have been used to demonstrate sorting and/or manipulation of biological molecules [23–29] tweezers. OT based on plasmonic substrates have also been developed to overcome the limitation imposed by free-space diffraction and to enhance the local optical intensity within the trap [12, 20, 21, 30–33]. A major advantage of plasmonic nanostructures is that the electromagnetic field enhancement is achieved over a broad range of incident laser wavelengths [34, 35] by modifying the geometry of the structure. Therefore, in combination with biology specimens, plasmonic OT (POT) could be employed for further investigation of the biological mechanisms behind the development of several diseases [31, 33]. Although several articles have reviewed various aspects of POT [20, 21, 31–33], new efforts with impact in various scientific fields have emerged. In this review, we will provide a detailed overview picture of the advances in POT based on metallic nanostructures, which have attracted considerable attention, as well as insights and perspectives for future applications. This review is structured as follows: first, we present a basic theoretical description of the physical origins of OT, which is useful to explore the underlying physical aspects of the momentum transfer process. Next, we review OT based on several plasmonic geometries, including a description of the different materials that have been used to enhance the optical forces in the nanoscale regime. In parallel, applications of POT in life sciences, especially for the manipulation of biomolecules such as proteins or DNA, are discussed. Finally, we provide both a critical view of the research field by summarizing the main conclusions extracted from the state of the art and our perspective of future POT developments and potential applications.

## 2 Theoretical aspects of plasmonic optical tweezers

Historically, in the 17th century, the astronomer Kepler proposed that the sun's radiation pressure causes comet tails to always point away from it [17]. In 1873, Maxwell [19] theoretically suggested that light could exert radiation pressure and this was demonstrated experimentally nearly a century later, after the advent of lasers [18, 36, 37]. Practically, the radiation pressure can be understood as a consequence of the conservation of momentum during the scattering and absorption of the photons. Additionally, electromagnetic fields induce a polarization in dielectric materials which results in a gradient force pointing toward high-intensity field regions. Therefore, optical trapping is based on a tightly focused beam [38] capable of generating a strong gradient force which overcomes the radiation pressure and holds or manipulates particles at the focal point [38]. Optical trapping theory is based on two limiting cases [39]: (1) for particles whose size is larger than the wavelength of the laser employed as the source of trapping ($r \gg \lambda$) and (2) for particles whose size is smaller than the wavelength of light ($r \ll \lambda$). In the first case, the condition of Mie scattering is satisfied and the trapping force can be calculated in terms of geometric ray optics [39]. In the latter case, Rayleigh scattering [11, 40] is dominant and the trapping force can be computed by treating the particle as a point dipole with an induced dipole momentum [40–42]. Moreover, for particles whose size is comparable with the trapping wavelength, neither the ray optic nor the point dipole approach is valid [39]. Therefore, a more complete electromagnetic theory analysis is required in order to supply an accurate description [39]. However, in this section we will discuss the nature of the optical force exerted on particles in the nanometer-sized regime.

## 2.1 Brief review of optical trapping principles in the dipole approximation

As mentioned above, the optical force is a consequence of the change in momentum carried by photons. However, the concept that an electromagnetic field carries momentum led to a long-standing debate known as the Abraham-Minkowski problem [5]. Consequently, the calculation of the optical force on a particle embedded in a viscous medium has followed different approaches, which can be derived by integrating the momentum flux over a closed surface surrounding the particle [5]. There are several articles which discuss, in detail, the optical force calculation based on a potential or the Maxwell stress tensor approach [5, 12, 19, 22, 36, 37, 41–44]. Here, by employing the dipole approximation to consider the optical force, the time-averaged force is divided into three components: the gradient force, the scattering force and the polarization gradient force, and is given by the following expression [43, 44]:

$$\langle F \rangle = \frac{1}{4}\text{Re}(\alpha_p)\nabla|E|^2 + \frac{\sigma(\alpha_p)}{2c}\text{Re}(E \times H^*) + \sigma(\alpha_p)c\nabla \times \left(\frac{\varepsilon_o}{4\omega i}E \times E^*\right) \quad (1)$$

where $\langle \ldots \rangle$ represents the time-averaged operation, Re denotes the real part, the asterisk denotes the complex conjugate, $\sigma(\alpha_p)$ is the extinction cross-section of the particle [43], $E$ and $H$ are the electric and magnetic fields, respectively, $c$ is the speed of light in vacuum, $\varepsilon_o$ is the vacuum dielectric permittivity, $\omega$ is the angular frequency of the optical field and $\alpha_p$ is the polarizability for small dielectric particles that is given by the following expression [45]:

$$\alpha_p = \frac{\alpha_o}{1 - i\alpha_o k_o^3/(6\pi\varepsilon_o)} \quad (2)$$

where $k$ is the wavenumber and $\alpha_o$ is the Clausius-Mossotti relation [45]:

$$\alpha_o = 4\pi\varepsilon_o r^3 \frac{\varepsilon_p - \varepsilon_m}{\varepsilon_p + 2\varepsilon_m} \quad (3)$$

Here, $\varepsilon_p$ and $\varepsilon_m$ are the dielectric constants of the particle and the surrounding medium, respectively, and $r$ is the radius of the particle. The gradient force, which is the first term in Eq. (1), attracts the particle along the direction of the increasing electric field intensity gradient [11, 12, 40]. The radiation force [13], the second term in Eq. (1), pushes the particle out of the trap. The third is the "spin curl force," which arises from the presence of the spatial polarization gradients [43]. This force is nonzero for a tightly focused beam in OT [44]. The balance between the above-mentioned forces produces an equilibrium point near the laser focus in which optical trapping in three dimensions can occur.

In the Rayleigh regime [40], the gradient force [11, 46] exerted on a trapped particle is proportional to the third power of the particle radius, that is, $r^3$, while the scattering force [11, 46] is proportional to the sixth power of the particle radius, that is, $r^6$. Hence, a small particle may easily escape from the trap due to its Brownian motion. Furthermore, for stable trapping of a nanoparticle, the trapping potential resulting from the optical forces should overcome the thermal energy [11]. The trapping potential energy exerted on a nanoparticle located at $r_o$, $U(r_o)$, is giving by the following expression [12]:

$$U(r_o) = -\int_{\infty}^{r_o} F(r) \cdot dr \quad (4)$$

and defines an important figure-of-merit for an optical trap. To achieve stable trapping, the depth of this potential well should be around 10 $k_B T$ (where $k_B$ is Boltzmann's constant and $T$ is the absolute temperature in the trap), in order to compensate for delocalization of the particle due to its Brownian motion [11]. Therefore, to confine a nanoparticle using the gradient force to counteract the destabilizing effects of Brownian motion requires either an increase in the light field intensity or that one works with highly polarizable particles. However, these requirements are neither feasible nor advisable for particles such as small biological specimens, for example, proteins or DNA strands, since they are temperature sensitive.

Thus, in order to overcome the obstacles that arise from using conventional OT, a novel technique based on principles of nano-optics was developed. This technique is frequently referred to as POT and provides us with several advantages. In 1997, it was proposed that a sharp metal tip illuminated by a laser beam could create sufficient optical potential to trap nanoparticles [23, 47]. This early work indicated that efficient optical trapping of the nanoparticle was achieved close to the surface of the metal tip. Specifically, in the presence of laser illumination, the free electron density on the surface of a metal structure can undergo collective oscillations that are called plasmons. Under certain conditions, the charge concentrations, due to geometrical features of the metal structure, create highly localized evanescent fields with an intensity that decays exponentially as moving away from the interface, thereby inducing strong gradient forces capable of increasing the precision of optical trapping. These fields due to localized surface plasmons can increase the confinement and the depth of the trapping potential, making the trapping more stable in the nanometer regime. Two years later, Okamoto et al. proposed a POT using confined light transmitted through a sub-wavelength aperture in an opaque metallic film [48]. Based on these significant early works, the coming-of-age of POT happened a decade later when small particles were trapped using low incident laser powers [20, 21, 31, 34], opening new roads for precisely trapping at the nanometer scale.

In another direction, a number of experiments using nanophotonic cavities, such as nanoapertures or photonic crystal cavities, demonstrated a new trapping mechanism known as self-induced back-action (SIBA) in which the trapped particle has a dynamic role on the trapping mechanism itself [49]. Generally, SIBA has been extensively studied and reported in the last decade for dielectric nanoparticles as well as plasmonic nanoparticles [49–53]. Here, the role of the SIBA mechanism for a plasmonic nanoaperture will be discussed briefly in the following sub-section [49].

## 2.2 Brief theory of SIBA trapping

A key component of SIBA trapping is that the particle itself plays an active role in the trapping mechanism by modifying the optical transmission signal through a sub-wavelength hole [49]. Generally, the origin of the SIBA effect arises from the sensitivity of the resonance to local refractive index changes. Specifically, Bethe first studied the diffraction of light transmitted through a tiny circular hole in an infinitely thin and perfect metallic film [54]. He showed that the optical transmission was proportional to the fourth power of the ratio of the hole radius to the light wavelength (i.e. $(r/\lambda)^4$). In other words, Bethe stated that light transmission through a hole, with a size smaller than the wavelength of light, in a metal film is very low. However, his predictions have been challenged by the discovery of the extraordinary optical transmission effect, where the transmission of light through a sub-wavelength hole is increased when the hole is arranged in a certain way [55]. Specifically, the transmitting light through a hole, in a real metal film, couples into surface plasmon polariton (SPP) modes, which penetrate into the hole wall effectively increasing the hole size. Consequently, the SPP modes on two opposite sides of the hole have a stronger coupling, giving a higher effective refractive index and, therefore, an increase in the optical transmission signal. In a similar way, by adding a particle into a nanohole, the effective refractive index increases and the particle makes the hole effectively larger due to the dielectric loading effect. Hence, an increase in the transmission signal by a factor related to the particle refractive index is obtained. Thus, the effective optical trapping achieved due to the nanohole structure evokes the strong fourth power scaling of Bethe's theory to overcome the third power polarizability of Rayleigh scattering. The benefit of the SIBA configuration over other POT approaches is the absence of plasmon resonance requirements.

## 2.3 Advantages and disadvantages of POT

Following the theoretical prediction of POT, several research groups experimentally demonstrated various designs of near-field techniques which paved a way for novel nanoscience and nanotechnology applications. Near-field trapping provides significant benefits compared to the far-field OT. Specifically, for POT, the trapping laser power, which is essential to achieve stable trapping in the Rayleigh regime, can be reduced by the electromagnetic enhancement of the SPP modes. This means that the POT can potentially be combined with applications such as the acceleration of photochemical reactions [56] or sensitive detection of biological entities [57, 58]. Moreover, it is well known that the diffraction limit of laser light places a lower limit on the size to which light can be focused. By employing the POT configuration, the motion of the trapped nanoparticle is confined in the plasmonic region, which is much smaller than the diffraction-limited area of the laser light, thereby providing a stable trap for particle diameters of 12 nm [59].

Although the POT allows for an increase to the local intensity by condensing light down to a smaller cross-sectional area, a fundamental concern is the thermal effect generated as a result of the frequency-dependent absorption of the laser light by the metallic nanostructures. The thermal convection effects have been shown to alter particle dynamics, giving rise to trapping phases in POT applications [60]. Such photothermal effects have been studied in POT implementations [60–62]. Ploschner et al. theoretically studied the optical forces exerted on glass nanobeads in the proximity of a gold nanoantenna and noted that the localization of the particles is due to heating effects, which probably dominate over optical forces [63]. Verschueren et al. demonstrated a temperature increase of 3.6 °C at metallic nanopore when illuminated with an incident power of 7.5 mW [64]. Xu et al. predicted a maximum temperature increase of 6 °C for double nanoaperture at an incident laser intensity of 6.67 mW/µm$^2$ [65]. More recently, Jiang et al. combined fluorescence spectroscopy with OT in order to locally measure the temperature of single and double nanoapertures [66]. The authors showed that the temperature can increase by around 10 °C at 2 mW/µm$^2$ incident laser intensity for a double nanoaperture and 20 °C under 5 mW/µm$^2$ illumination for a single nanoaperture [66]. Notably, the majority of POT experiments are performed using a base wafer material of either fused silica or silicon. The thermal conductivity of fused silica is almost 2.5 times greater than that of water and silicon is close to 250 times greater. Moreover, Roxworthy et al. performed theoretical calculations of the plasmon-induced convection flow above an array of gold bowtie nanostructures on an indium tin oxide (ITO) thin layer and pointed out that the thermally conductive ITO can distribute the thermal energy efficiently compared to the same plasmonic geometry on a less thermally conductive substrate layer [60]. Therefore, in order to suppress the photothermal effect, the following approaches have been proposed: the fabrication of plasmonic nanostructures on a heat sink [67, 68] and the decrease of the number of the plasmonic nanostructures within the illuminated area [61]. Consequently, thermal energy generated by absorption of optical energy in the water solution surrounding a near-field trap can be better dissipated through the underlying substrate. In spite of this, it is also essential to note how the optical thermal effect can be utilized to facilitate temperature-related applications such as biosensing [69–72]. For example, metallic bowtie nanoantenna arrays [73], single nanoantenna assisted with an applied electric field [74], a thermal absorption medium [75] and continuous gold films [76] have achieved the manipulation and transportation of small particles through the assistance of plasmon-induced thermal convection. Therefore, the thermophoretic force is an essential factor and should be taken into account in any exploration of the POT mechanism.

## 3 Plasmonic nanostructures

Plasmonic nanostructures have recently drawn much attention due to their ability to overcome the diffraction limit of dielectric configurations and, as such, have been used to manipulate nanosized particles. They can provide strong, sub-wavelength energy confinement with a high optical gradient at the interface between the plasmonic and dielectric layers [77], making them suitable for novel applications in optical communications [78, 79], optical imaging [80], energy harvesting [81], nanoelectronics [82, 83], sensing [84–86] and optical trapping [20, 21, 31–33, 87]. Several metallic nanostructure designs have been proposed for near-field OT applications. In this section we highlight the plasmonic nanostructures that are relevant for optical trapping and briefly discuss their applications, mainly in biophysics and medical science, but also some recent advances down to the nanoscale regime. In Table 1, we summarize the important parameters of these POT for micro-/nanosized particle manipulation.

### 3.1 POT on a homogeneous metallic film

The first experimental implementation of a POT was demonstrated on a homogeneous gold-dielectric interface [88, 110]. In 2006, Garcès-Chávez et al. performed experiments on a 40-nm gold thin film by employing the Kretschmann-Raether configuration [110]. They observed the self-assembly of a large number of 5 µm silica beads using a combination of optical and thermophoretic forces. They showed that by controlling the excitation of the surface plasmon, a hexagonal, close-packed crystalline arrangement of an array of 2800 beads was achieved [110]. The same year, Volpe et al., by employing a photonic force microscope, measured the radiation force as a function of distance from a 40-nm gold layer, for various diameters of trapped particles. They observed that the magnitude of the radiation force decays but did not follow the exponential tail of the plasmon field, on moving away from the gold substrate [88]. Moreover, they reported the first observation of momentum transfer from an SPP to a single dielectric particle of diameter 4.5 µm and measured an optical force enhancement of up to 40× at resonance compared to that for non-resonant conditions [88]. With a similar configuration scheme, a POT based on a 45-nm gold thin layer was developed to trap and manipulate gold particles of 0.5–2.2 µm diameter when the surface plasmon was excited by a radially polarized cylindrical vector beam [111]. The authors noted that if a high numerical aperture objective lens is used, the SPP wavefront can be excited from the converging trapping laser beam in the homogeneous metallic layer and propagates toward the center of the focal spot [111]. They also determined that the total optical force originates from the coupling between a greatly enhanced SSP field and gold nanoparticles [111].

**Table 1:** Trapping parameters for various POT configurations.

| Plasmonic nanostructure | | | | Light source | | | Trapped nanoparticle | | Exp./sim. |
|---|---|---|---|---|---|---|---|---|---|
| Geometry | Enhancement factor $\eta=\|E_{max}\|^2/\|E_i\|^2$ | Trap stiffness | Dimensionless trapping efficiency $Q$ | Wavelength (nm) | Intensity (mW/μm²) | Power (mW) | Material | Diameter (nm) | |
| Gold film | | 1.1 pN/μm | | 632.8 | | 18 | Polystyrene [88] | 4500 | Exp. |
| | | 1.3 pN/μm | | 632.8 | | 18 | Polystyrene [88] | 2000 | Exp. |
| | | 1.9 pN/μm | | 632.8 | | 18 | Polystyrene [88] | 600 | Exp. |
| Gold microdisks | | $k_x$ = 27 fN/μm | | 710 | 2.5×10⁻³ | 250 | Polystyrene [89] | 4880 | Sim. |
| | | $k_y$ = 22 fN/μm | | | | | | | Sim. |
| | | $k_x$ = 26 fN/μm | | 710 | 2.5×10⁻³ | 250 | Polystyrene [89] | 3550 | Sim. |
| | | $k_y$ = 17 fN/μm | | | | | | | Sim. |
| Nanoantenna | 20 | $k_r$ = 19 pN/μm | | 974 | 10 | | Polystyrene [67] | 110 | Exp. |
| | 20 | $k_r$ = 19 pN/μm | | 974 | 5 | 2.5 | Polystyrene [67] | 200 | Exp. |
| | | $k_t$ = 4.8 pN/μm | | | | | | | |
| Double nanoantenna | 30 | 13 pN/μm | 0.1 | 1064 | 5×10³ | 155 | Polystyrene [34] | 200 | Exp. |
| | 30 | | 14 | 1064 | | 532 | Polystyrene [34] | 6000 | Exp. |
| | 30 | | 1.6 | 1064 | | 440 | Polystyrene [34] | 1000 | Exp. |
| | | | 45 | 1064 | 100 | 14 | Bubble [90] | 1400 | Exp. |
| Four nanodisks | 20 | | | 980 | 2.8 | 3.5 | Polystyrene [91] | 100 | Exp. |
| | 20 | | | 980 | 2.8 | 4.75 | Polystyrene [91] | 500 | Exp. |
| Nanoaperture | 7 | $k_x$ = 6.6 pN/nm | | 1064 | 2 | 1.9 | Polystyrene [50] | 50 | Sim. |
| | 7 | $k_y$ = 9.3 pN/nm | | 1064 | | 0.7 | Polystyrene [50] | 100 | Sim. |
| Double nanoaperture | | | | 975 | 5 | <10 | Polystyrene [59] | 20 | Exp. |
| | | 0.2 fN/nm · mW | | 975 | 7.2 | 7.2 | Silica [59] | 12 | Exp. |
| | | | | 820 | | 2 | Polystyrene [92] | 20 | Exp. |
| | | | | 820 | 3.5 | 3.5 | Bovine serum albumin [93] | 6.8 | Exp. |
| | 25 | 1.02 fN/nm | | 855 | | 8.25 | Magnetite [94] | 30 | Exp. |
| | | 0.38 fN/nm | | 855 | | | Polystyrene [94] | 30 | Exp. |
| | 16.3 | $k_x$ = 0.2625 fN/nm | | 1064 | 6.6 | | Polystyrene [65] | 20 | Exp. |
| | | $k_y$ = 0.0801 fN/nm | | | | | | | |
| Nanoaperture arrays | | 1.07 pN/μm · mW | | 980 | 0.6 | | Polystyrene [95] | 1000 | Exp. |
| | | 0.25 pN/μm · mW | | 980 | 1.5 | | Polystyrene [95] | 500 | Exp. |
| | | 0.85 fN/nm · mW | | 980 | 0.51 | | Polystyrene [96] | 30 | Exp. |
| Coaxial nanoaperture | 25 | $k_x$ = 1.05 fN/nm · mW | | 818 | | 67 | Polystyrene [97] | 10 | Sim. |
| | | $k_y$ = 0.19 fN/nm · mW | | | | | | | |

Table 1 (continued)

| Plasmonic nanostructure | | | Light source | | | Trapped nanoparticle | | Exp./sim. |
|---|---|---|---|---|---|---|---|---|
| Geometry | Enhancement factor $\eta=\|E_{max}\|^2/\|E_i\|^2$ | Trap stiffness | Dimensionless trapping efficiency Q | Wavelength (nm) | Intensity (mW/μm²) | Power (mW) | Material | Diameter (nm) | |
| Double nanorods | 10 | | | 800 | 0.01 | 300 | E. coli [98] | 2000 | Exp. |
| | 10 | | | 800 | 0.01 | 300 | Polystyrene [98] | 200 | Exp. |
| | 30 | | 0.4 | 808 | 2 | 800 | Gold [61] | 10 | Exp. |
| Double nanopyramids | 10⁴ | | | 808 | 0.05 | <1.0 | Quantum dot [99] | 10 | Exp. |
| | 10⁴ | | | 808 | 0.1 | <1.0 | Polymer chain [100] | 100 | Exp. |
| | | | | 770/808 | 0.7 | | λ-DNA [101] | (48.5 kbp) | Exp. |
| Double nanoblocks | 6300 | $k_x$ = 2.3 fN/nm $k_y$ = 2.0 fN/nm | | 800 | 7.5×10⁻³ | 0.015 | Polystyrene [102] | 350 | Exp. |
| | 398 | | | 1064 | 0.6 | | Polystyrene [103] | 100 | Exp. |
| Nanobowtie | 300 | 14 pN/μm·mW | | 800 | 20 | 0.05 | Polystyrene [104] | 1200 | Exp. |
| | | | | 800 | | 0.065 | Silver [104] | 80 | Exp. |
| Bowtie nanoaperture | 280 | 2.4 fN/nm | | 1064 | 1 | 10 | Gold [53] | 60 | Exp. |
| | 100 | $k_x$ = 0.42 fN/nm·mW $k_y$ = 0.07 fN/nm·mW | | 1064 | 15.6 | | Quantum dot (scQD) [105] | 30 | Sim. |
| Double nanobowtie | | | 0.27 | 1064 | 1 | 10 | Gold [53] | 60 | Exp. |
| | | | 0.65 | 685 | 0.014 | 0.05 | Polystyrene [73] | 500 | Exp. |
| | | | 1.72 | 685 | 0.068 | 0.025 | Polystyrene [73] | 1000 | Exp. |
| | | | | 685 | 0.044 | 0.015 | Polystyrene [73] | 1500 | Exp. |
| Nanodiabolo | 260 | $k_r$ = 0.69 pN/nm·W 0.34 pN/nm·W | | 980 | 3.57 | 5.5 | Polystyrene [106] | 300 | Exp. |
| | | | | 980 | 6.1 | | Polystyrene [106] | 300 | Sim. |
| Silver-coated black Si | | | 0.054 | 1064 | | | Polystyrene [107] | 900 | Exp. |
| Gold-coated black Si | | | 0.117 | 1070 | | | Polystyrene [108] | 400 | Exp. |
| | | | 0.203 | 1000 | | 24.5 | Polystyrene [109] | 400 | Exp. |

## 3.2 POT on metallic disks

Although the thin metallic layer illuminated by a laser beam creates an optical trapping field, stable optical trapping at a specific location on the metallic surface requires a confined trapping well, which can be achieved through metal patterning. A finite, gold, micron-sized disk pattern was designed at the surface of a glass substrate in order to manipulate micron-sized polystyrene particles [35]. In this experiment, the patterned surface was illuminated through a hemi-cylindrical glass prism by an unfocused linearly $p$-polarized laser beam at 785 nm. The illuminated spot was smaller than the gold disk dimensions, in order to achieve parallel trapping of a large number of particles under the same illumination conditions. Moreover, the authors observed that the trapping event lasted for several hours when using a laser intensity around $5 \times 10^7$ W/m$^2$, which is considerably lower than that required for conventional OT. A year later, the same research group [89], employing a similar POT setup, investigated the confinement and the stiffness of the trapping process, revealing stable trapping of 4.88 μm polystyrene beads, with a trap stiffness in the range of a few tens of fN μm$^{-1}$. Although the micron-sized disk geometry enables particles as small as 1 μm to be trapped, by down-scaling the disk diameter, the trapping process for smaller particle diameters is still limited [89]. This observation led to the design and fabrication of alternative plasmonic geometries in which much higher control of the plasmonic fields could be achieved.

A few years later, Chen et al. studied the trapping behavior of 100 and 500 nm polystyrene beads in a plasmon-enhanced, two-dimensional optical lattice [91]. The authors fabricated a square lattice with a period of 1 μm in which each primitive cell consisted of four gold nanodisks of 200 nm diameter with the thickness of 40 nm (Figure 2A) [91]. These plasmonic substrates were used to efficiently guide and arrange nanoparticles with the intention to investigate coherent interactions and explore novel physics issues [91]. Another POT approach was based on silver nanodisks with a graded diameter for delivering the gold-trapped particle in a controllable way [113]. In this work, the authors showed that the position of the hotspots can be changed by changing the incident wavelength and rotating the polarization state [113]. This approach offers an alternative way for particle transportation in the nanoscale regime [113].

## 3.3 POT on pillar/antenna nanostructures

Among the most interesting plasmonic geometries, metallic nanopillars or nanoantennas have the capability to concentrate light into a very localized and intense hotspot. Regarding this benefit, they can create optical traps orders of magnitude smaller than those achievable through the micron-sized disk approach. Thus, in this direction, Sidorov et al. reported the first experimental realization of a POT based on gold periodic nanopillars [90]. In their experiments, microbubbles were created, trapped and manipulated by a 1064-nm laser beam in an immersion oil medium. The authors showed sub-micron precision of microbubble position as well as a high trapping efficiency ($Q \sim 50$ where $Q$ is the dimensionless trapping efficiency which represents the fraction of incident laser power utilized to exert a force [34]) [90]. A year later, Grigorenko et al. reported the trapping mechanism of particles near gold periodic nanopillars [34, 114]. A 1064-nm focused laser beam was used to trap a 200-nm polystyrene particle in an immersion oil medium [34]. The authors observed that the particle could be moved between adjacent nanopillars by translating the laser beam, thereby allowing for trapped particles to be discretely positioned. They also measured a trapping force of 2 nN, which is an order of magnitude higher than that obtained with conventional OT [34]. They calculated a high trapping quality factor of 1.6 near the plasmonic substrate for 1-μm-diameter particles [34]. It should be noted that, in this experiment, the trapped particles were suspended in an oil medium and high light intensity was required in comparison with similar POT systems [34].

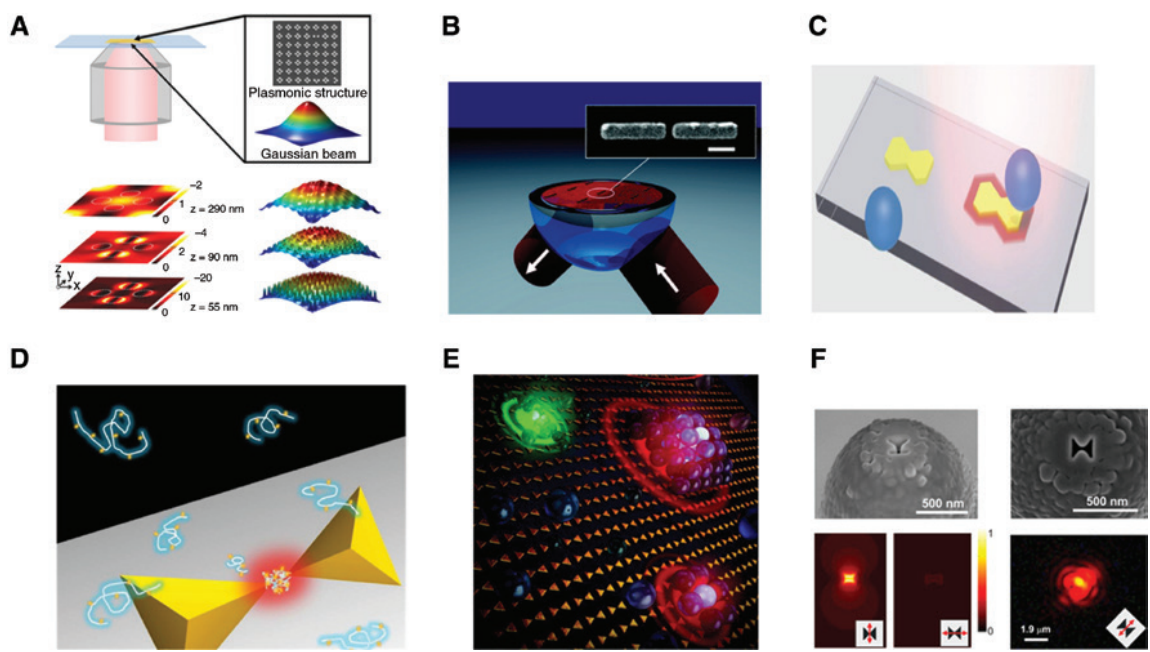

**Figure 2:** Plasmonic optical tweezers designs.
(A) Four nanodisk POT, reprinted with permission from [91], Copyright 2013 American Chemical Society. (B) Nanoantenna POT, reprinted with permission from [98], Copyright 2009 American Chemical Society. (C) Nanodiabolo POT, reprinted with permission from [106], Copyright 2011 Nature Publishing Group. (D) Nanopyramids POT, reprinted with permission from [100], Copyright 2012 American Chemical Society. (E) Bowtie arrays POT, reprinted with permission from [73], Copyright 2012 American Chemical Society. (F) Fiber nanotweezers reprinted with permission from [112], Copyright 2014 The Optical Society.

In addition to optical forces, coupling to SPPs leads to local heating of the metal and heat dissipation to the surrounding fluid. In order to overcome this issue, a template-stripped plasmonic nanopillar was demonstrated to trap and rotate nanoparticles of 110 nm diameter [62, 67]. This POT configuration comprises a gold nanopillar on a copper layer formed on a silicon substrate. Due to the high thermal conductivities of the above-mentioned materials, the heat generated by the plasmonic field is conducted into the silicon substrate rather than into the water, therefore producing less water heating than in previous design schemes; hence higher laser powers can be used [67]. The authors showed that the nanopillar can convert the angular momentum of the incident laser beam to the orbital momentum of the inspected nanoparticle [62]. This observation paved the way for nanoscale motor functionality [67].

A POT approach based on a periodic array of nanopillars, exploiting the localized and extended surface plasmon modes characteristics, was proposed for simultaneous biosensing and optical trapping applications [115]. In this work, the authors noted optical forces of 350 pN μm$^{-1}$ W$^{-1}$ when the trapped particle was positioned close to a nanopillar. They also mentioned that directional control of the optical forces was achieved by controlling the incident light polarization [115]. A remarkable improvement of the POT efficiency has been demonstrated with a two-dimensional dipole gold antenna array, separated by a nanoscale dielectric gap [61]. The above-mentioned geometry provides a strong light spot within the narrow gap on resonance. The authors noted that the optical force increased by a factor of 8 when the particle size was changed from 10 to 20 nm [61].

Stable trapping of living matter, such as *E. coli* bacterium, has been observed for several hours with a dipole gold nanoantenna [98] (Figure 2B), with an input power of 300 mW, corresponding to an optical intensity of 10$^7$ W/m$^2$ near the plasmonic cavity; this is almost an order of magnitude lower than the damage threshold of bacterium [116]. The experimental implementation of plasmonic optical vortex trapping was demonstrated by an array of gold *diabolo* nanoantennas with a period of 2 μm, on a 15-nm gold thin layer (Figure 2C) [106]. The authors demonstrated the trapping of 300-nm-diameter fluorescent polystyrene beads suspended in deionized water, at the boundary of the plasmonic diabolo nanoantenna and measured a large radial trap stiffness equal to 0.69 pN nm$^{-1}$ W$^{-1}$, at the position of the minimum potential [106]. Moreover, they performed plasmonic trapping experiments using silica nanoparticles immersed in oil and observed that particles were trapped at the local minimum of the electric field intensity, that is, at the center of the diabolo nanoantenna [106]. This work exhibits the ability to effectively trap nanoparticles with a refractive index lower than the surrounding medium [106].

### 3.4 POT on pyramidal nanostructure

By employing the angle-resolved nanosphere lithography method, gold nanopyramid arrays on a glass substrate were fabricated for quantum dot optical trapping with laser intensities in the range of 0.5–10 kW/cm$^2$ [99]. In this experiment, the height of each nanopyramid and the distance between adjacent pyramids were evaluated to be 30 and 140 nm, respectively [99]. The authors mentioned that the optical trapping at nanovalleys, located between adjacent pyramids, induced photoluminescence quenching and enhancement [99]. Two years later, gold nanopyramidal arrays, which have an extinction band at wavelengths longer than 650 nm corresponding to a gap-mode localized surface plasmon, were fabricated in order to enable molecular manipulation (Figure 2D) [100]. In this work, near-infrared laser light at 808 nm was used to excite the surface plasmon, resulting in an enhanced electromagnetic field in close proximity to the gap between the nanopyramids [100]. Thereafter, the authors compared the trapping efficiency for two different kinds of dye-doped polystyrene nanospheres. One was resonant with the trapping laser light and the other non-resonant [100]. They concluded that the POT's performance under the resonant condition was at least five times better than that under the non-resonant condition, thereby providing an alternative approach for the efficient trapping of small molecules [100, 117].

Furthermore, the same research group employed the nanopyramids as a plasmonic substrate to optically trap fluorescent-labeled $\lambda$-DNA with a femtosecond (fs) laser trapping beam [101]. The authors observed that, on plasmon excitation with the fs laser, a microassembly was formed on the plasmonic area – the particles were released after stopping the laser irradiation [101]. They also claimed that POT of DNA with an fs laser will pave the way for efficient trapping methods for other biomolecules [101]. Recently, a nanopyramid substrate was fabricated to provide an alternative method for the detection of organic molecules dissolved in aqueous solution [118]. In this work, the molecules were extracted into a poly(N-isopropylacrylamide) microassembly formed by plasmonic optical trapping [118]. The fluorescence enhancement factor was measured to be up to 140, indicating that the molecules were detected at the $10^{-9}$ mol/l level [118]. These results strongly suggest that this POT implementation will provide a novel analytical method to detect a variety of organic molecules.

## 3.5 POT on bowtie nanostructures

Bowtie nanostructures [119–121] have garnered increasing interest in the nanophotonics community. Initially, the bowtie structure was reported in the microwave regime of the electromagnetic spectrum [120]; since then several bowtie configurations have been studied at mid-infrared wavelengths near to 10 $\mu$m [121]. The bowtie nanostructure consists of two metallic, symmetric triangular structures, facing tip-to-tip, but separated by a narrow gap. Theoretically, it has been shown that bowtie nanostructures have a higher sensitivity than a pair of nanoantennas [122]. In fact, the bowtie geometry can tightly focus light in a single spot inside the narrow gap. Even though the nanoantenna provides stronger energy peaks than the bowtie configuration, the latter are characterized by multiple optical traps at the external edges of the metal arms. Therefore, the bowtie configuration provides a strong advantage due to the position of the optical trap since it is well defined, resulting in a strong restoring force [53].

Exploiting this benefit, in 2012, Roxworthy et al. fabricated arrays of two equilateral gold triangles, each with a 120 nm tip-to-base height and separated by a 20 nm gap for trapping, stacking and sorting of sub-micron to micron-sized dielectric particles using a low trapping power of approximately less than 1 mW (Figure 2E) [73]. In this work, the authors measured a high trapping efficiency equal to 0.27 for a 500-nm-diameter trapped particle at 25 nm above the plasmonic substrate [73]. Moreover, the same research group demonstrated a POT based on an array of gold bowtie nanostructures with a 100 fs pulsed Ti:sapphire laser (80 MHz repetition rate, $\lambda=800$ nm) as an excitation source [88]. The authors compared the fluorescence intensity of a trapped particle of 600 nm radius when using the fs laser with the corresponding one for a continuous wave (cw) laser [104]. They noted that the fs laser excitation led to an increase in fluorescence intensity owing to the increase in radiation force [104]. They also measured the trap stiffness of the POT with an fs laser and found that it was equal to 14 pN $\mu$m$^{-1}$ mW$^{-1}$, two times higher than that of a POT with a cw laser source and five times more than that of a conventional OT with an fs laser source [104]. By applying the fs POT, they optically trapped silver nanoparticles of 80 nm diameter in the nanogap and observed adhesion of these particles to plasmonic active sites [104]. This effect had never been observed before by employing a POT with cw lasers [104].

Moreover, Lu et al. [123] proposed a theoretical investigation of the tunability properties of the potential well width, depth and stiffness of the POT based on bowtie nanostructures versus trapping laser wavelength. This research group demonstrated that optical trapping with low laser intensity can be realized with a 1280 nm laser wavelength, while trapping with high precision can be realized with a 1300 nm laser wavelength for the bowtie configuration [123]. Recent work, based on a bowtie geometry in a silver film, was demonstrated to isolate quantum dots and excite them with two-photon luminescence due to strong field enhancement inside the bowtie nanoaperture [124]. The authors measured the trap stiffness a 20-nm silica-coated quantum dot experiences inside the nanoaperture as being equal to 0.42 and 0.07 fN nm$^{-1}$ mW$^{-1}$ for the $x$ and $z$ directions, respectively. These values are nearly two orders of magnitude larger than in the case of free-space trapping [124].

Additionally, Mestres et al. fabricated bowtie nanoapertures milled by a focused ion beam in a 100-nm gold thin layer to optically trap gold nanoparticles of 60 nm diameter in order to investigate the optomechanical interaction of the plasmonic nanocavity [53]. They concluded that as the optical trapping power is kept low, the SIBA effect gets stronger until it becomes the main trapping mechanism [53]. This experiment provided a means to identify the optimum conditions that maximize the trapping efficiency under low trapping laser power and improve the trapping performance for the manipulation of photodamage-sensitive particles [53].

Recently, the ionic-current and optical transmission detection of single molecule λ-DNA translocations through a bowtie-shaped structure were reported [64, 72]. The bowtie structure was fabricated in a 100-nm gold thin film and placed on a 20-nm thin free-standing silicon-nitride membrane [64]. The authors demonstrated a label-free optical technique using plasmonic bowtie nanostructures that enables the conformation of translocation biomolecules or the study of thermophoretical polymeric translocations, while omitting an electrical bias [64, 72].

### 3.5.1 Bowtie nanostructures onto optical waveguides

A few years later, a different configuration of bowtie nanostructures for optical trapping was proposed [112]. The configuration was based on the integration of metallic nanobowties with optical waveguides. By coupling laser light with a wavelength of 1500 nm into the waveguide through a fiber, the metallic bowtie nanostructures generated highly concentrated resonant fields and induced strong optical trapping forces [112]. The authors showed that the theoretical values of the trapping forces were equal to 652 pN/W on particles with diameters as small as 20 nm [112]. Another design [125] involves the integration of the bowtie apertures on fiber tips to implement 20 nm polystyrene bead trapping and 50 nm bead manipulation with local intensities within the trap as small as $10^9$ W/m$^2$. This configuration has the potential to trap and translocate a particle to various locations with high accuracy [125]. The same year, Eter et al. [105] proposed the concept of a fiber-integrated optical nanotweezers based on a single bowtie-aperture nanoantenna fabricated at the apex of a 150-nm aluminum-coated near-field scanning optical microscope fiber tip (Figure 2F). They demonstrated 3D optical trapping of 500 nm latex beads with 1 mW input trapping power and estimated the magnitude of the optical forces induced by the bowtie nanoaperture on the fiber tip as being one thousand times larger than those obtained with a circular aperture with the same active area [105].

## 3.6 POT on aperture nanostructures

An alternative approach for POT involves the exploitation of the trap particle interaction to accomplish automatic feedback control. In 2009, a circular nanoaperture in a 100-nm gold thin film was used to trap a 50 nm polystyrene sphere with low optical trapping power, for example, less than 2 mW or local intensities lower than $10^9$ W/m$^2$, drastically reducing the likelihood of optical damage (Figure 3A) [50]. The authors employed the strong influence of the nanoaperture transmission due to particle dielectric loading to achieve efficient trapping [50]. Moreover, this work indicates that the proposed POT could be viewed as sensors in which the trapped particle has a strong influence on the local electric field to achieve efficient trapping [31]. Although the proposed POT was used to stably trap 50-nm-diameter particles with low trapping laser power [50], extending optical trapping toward even smaller particles requires new nanoaperture designs. Trapping of 22-nm-diameter polystyrene beads was achieved using a rectangular plasmonic nanopore (Figure 3B) [126]. The authors showed that this configuration has the ability to tune the propagating gap plasmons by adjusting the geometrical aspect ratio [126]. Moreover, they mentioned that this geometry can generate a strong field intensity gradient enabling trapping and enhancement of the far-field light transmission, thus facilitating monitoring of the trapping events despite the small aperture area [126]. Single- and double-bead trapping were observed, with the second bead experiencing a force field resulting from the presence of the first bead [126].

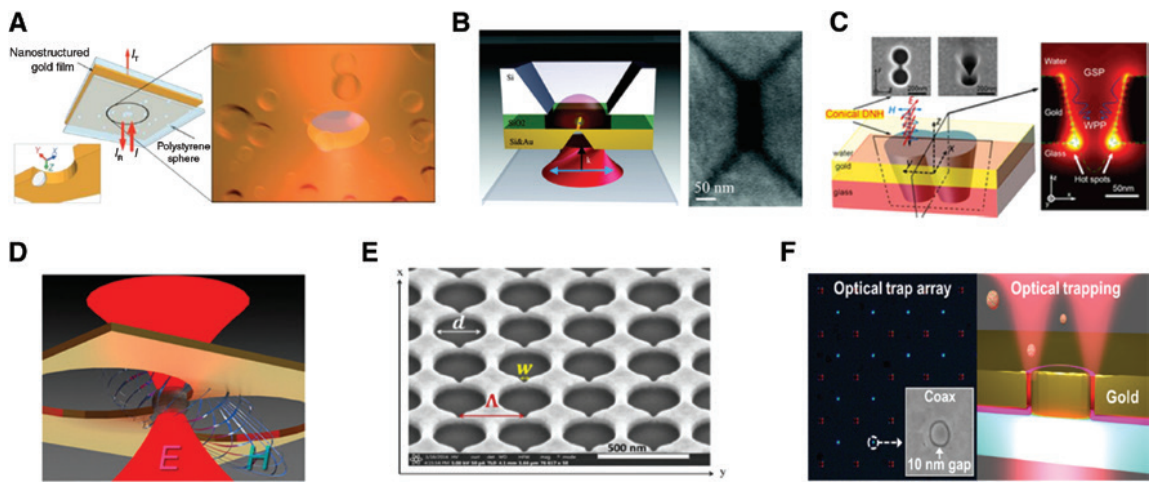

**Figure 3:** Nanoaperture Plasmonic Tweezers Configurations.
(A) Circular nanoaperture OT, reprinted with permission from [50], Copyright 2009 Nature Publishing Group. (B) Rectangular aperture OT, reprinted with permission from [126], Copyright 2012 American Chemical Society. (C) Double nanoholes with tapered cups OT, reprinted with permission from [127], Copyright 2017 American Chemical Society. (D) Double nanohole OT, reprinted with permission from [94], Copyright 2016 American Chemical Society. (E) Annular nanoapertures OT, reprinted with permission from [96], Copyright 2018 The Optical Society. (F) Coaxial nanoaperture POT, reprinted with permission from [128], Copyright 2018 American Chemical Society.

Double nanoapertures in a gold film, which strongly confine the electromagnetic field [31, 59, 92, 93, 129–139] at the cusps where holes overlap, have been employed for trapping a 12 nm silica sphere [59], for trapping and unfolding a single protein [93], for unzipping a single DNA-hairpin [136], for studying the interaction between a single protein with small molecule [131, 132, 140] and for characterizing the molecular weight of a single protein [141]. Double nanoapertures have the ability to trap small particles more easily than larger particles due to the sensitivity to the gap size between the two nanoapertures [31]. Specifically, Gordon's research group has demonstrated trapping of a single BSA protein in a double nanoaperture by employing at least 3.5 mW laser trapping power and observed the transition between protein folding and unfolding configurations [93]. Two years later, the same research group distinguished the bound and unbound forms of a single protein [136], the binding of zipping and unzipping of single DNA [31] and the single protein binding kinetics of a single human albumin [130]. Kontala et al. used a double nanoaperture POT with two trapping lasers beating to excite the vibration modes of single-stranded DNA fragments in the 10–100 GHz frequency range [137]. This approach established the ability to characterize small DNA strands with resolution of a few base pairs and has the potential to be extended further for exact base sequence determination [137]. Recently, Hacohen et al. developed a double nanohole of 90 nm radius and 35 nm line slot based on a 100-nm gold layer in order to analyze the protein composition of unpurified heterogeneous medium, that is, egg white [134]. By employing autocorrelation analysis, the authors showed a 2/3 dependence of the time constant with the molecular mass [134, 139]. The double nanoaperture structure has been used extensively to trap and sense biological and dielectric particles.

Unlike the cylindrical double nanoaperture approach, a conical wedge, double nanohole structure was developed to improve the trapping of 11 nm radius polystyrene particles [127]. The authors claimed that this design leads to two-dimensional nanofocusing of the gap surface plasmons and couples them to the wedge plasmon modes, thus creating a hotspot required for trapping (Figure 3C) [127]. Experimental results showed that due to the sensitivity of the gap size between the double nanoaperture, the small particle could be trapped more easily than a larger particle [136]. The fabrication of those double nanoholes can be expensive with low throughput. Thus, a method based on template stripping was proposed to pattern double nanoholes as small as 7 nm [141]. The trapping capability of the new plasmonic double nanoapertures prepared by template stripping was confirmed by optically manipulating the protein streptavidin [141]. The authors showed that there are no differences in the trapping performance with respect to traps prepared by direct focused ion beam milling of gold films, in terms of the time to trap, step size observed in the transmission signal, trapping duration and relative response to varied trap sizes [141]. The precise control of the position and the orientation of elongated particles led the way toward new applications in many areas of science.

Aporvari et al. [142] studied theoretically the optical trapping and control of a single dielectric nanowire, placed at a small distance above a single nanoaperture of radius 155 nm in a 100-nm gold layer. They demonstrated that the nanoaperture could efficiently trap dielectric nanowires with low power and control their orientation and movement through the polarization of the input beam [142]. Additionally, Xu et al. studied paramagnetic nanoparticles of 30 nm diameter, combining a POT with a double nanoaperture geometry and magnetic field arrangement (Figure 3D) [94]. In this study, a diode laser at 855 nm with a maximum trapping power of 15 mW was focused onto a 200-nm-diameter gold nanoaperture with a 35 nm tip [94]. The authors obtained a size distribution of magnetite comparable with both supplier specifications and scanning electron imaging measurements, indicating that their POT scheme could be a powerful tool for isolating desired particles [94]. Kwak et al. demonstrated trapping of 200 nm fluorescent polystyrene beads using a nanoaperture of 500 nm diameter in a 250-nm gold film [143]. Another interesting approach is the direct particle tracking observation of 20-nm-diameter polystyrene nansospheres suspended in deionized water by a double nanoaperture in a gold film [65]. The authors showed that the position distribution of the trapped particles in the plane of the film has an elliptical shape [65].

### 3.6.1 Nanoapertures onto optical waveguides

Integrating nanoapertures onto the facet of an optical fiber has the potential to open up new perspectives for chemical and biological sensing studies [144]. Therefore, optical trapping of a single polystyrene sphere was demonstrated at the cleaved end of a fiber with a double nanoaperture in a gold layer and without any microscope optics [144]. The integration of the nanohole on the tip of the fiber was developed by depositing 5 nm of titanium and 80 nm of gold followed by a 20 nm angled sputter deposition of gold/palladium to the fiber tips [144]. In this work, the optical transmission increased by 15% for the 40-nm-diameter sphere and 2% for the 20-nm-diameter sphere trapping events [144]. This POT approach can be used to replace the typical optical trapping setups that require complicated free-space optics and frequent calibration and alignment.

### 3.6.2 Nanoapertures arrays

A configuration of a plasmonic nanoring array which provides tunability of the resonance frequency in the near infrared was demonstrated for optical manipulation and sorting of nanometer-sized particles [145]. Specifically, in this experiment, a gold nanoring array with a 50 nm connecting gap and outer ring diameter of 300 nm, as well as various inner diameter sizes, was fabricated by employing electron beam lithography [145]. The authors demonstrated the trapping performance of the plasmonic nanostructures by optically manipulating 100 nm polystyrene particles with a low incident trapping power of 1 mW [145]. Moreover, the authors claimed that by using the abrupt phase changes of the proposed plasmonic nanostructures one can improve the detection sensitivity by a factor of 10 compared to the corresponding excitation spectra method [71], opening the way for the development of a non-destructive and highly sensitive tool for single molecule detection [145]. Recently, Han et al. demonstrated optical trapping of micro- and nanoparticles using a POT based on arrays of annular nanoapertures (Figure 3E) [95, 96]. The authors measured a trap stiffnesses of 0.25 and 1.07 pN $\mu m^{-1}$ mW$^{-1}$ for 0.5- and 1-$\mu$m-diameter particles, respectively [95]. Moreover, the authors observed particle transportation across the plasmonic hotspots with low laser intensities [95]. They performed, for the first time, sequential single-nanoparticle trapping within specific trapping sites and they obtained a trap stiffness of 0.85 fN nm$^{-1}$ mW$^{-1}$ at a low incident laser intensity of around 0.51 mW/$\mu m^2$ for 30-nm-diameter polystyrene particles [96]. The authors noted that their POT design has the potential to be used in lab-on-a-chip devices for efficient particle trapping with high tunability of the Fano resonance wavelength [96].

### 3.6.3 Coaxial nanoapertures

The optical force and the trapping potential well depth were theoretically studied by employing a POT based on a coaxial plasmonic structure [97]. This structure consisted of a 150-nm silver slab with an embedded silica ring. The authors calculated the maximum restoring force and trap stiffness in the $x$ direction as 1.88 pN/100 mW and 1.05 fN nm$^{-1}$ mW$^{-1}$, respectively [97]. A few years later, the same research group designed an achiral coaxial metallic nanoaperture to selectively trap sub-10 nm particles based on their chirality [146]. The authors showed that S enantiomers with matched handedness of the incident light could be trapped within 20 nm above the nanoaperture, while the R enantiomers remained untrapped due to a positive trapping potential well depth at the same location [146]. They also noted that the total maximum transverse force for the S and R enantiomers was 1.23 and 0.35 pN/100 mW, respectively [146]. This coaxial nanoaperture design provides for selective trapping of small particles based on their chiralities and a method to study their interactions with other chiral components [146]. More recently, Yoo et al. demonstrated optical trapping of a 30 nm polystyrene particle and streptavidin molecules with a laser power at 4.5 mW by employing a 10-nm gap resonant coaxial nanoaperture in a gold film (Figure 3F) [128]. Specifically, the authors selected the first-order Fabry-Pérot mode of the coaxial nanoaperture due to the ease in tuning the resonance in the near infrared to stably trap protein molecules within 3 min in a reproducible manner [128]. They claimed that their POT design pushes the limit of nanotrapping technologies due to the sharp trapping potential well [128].

### 3.6.4 C-shaped nanoapertures

A trapped particle may also be transported by employing optical forces. Hansen et al. proposed a nano-optical conveyor design to independently control the movement of a particle from one trap to another [147]. The geometry is based on repeating linear structures of three district traps by C-shaped apertures [147]. Each element of the C-shaped aperture was separately addressable by their resonant wavelengths and polarizations [147]. The authors showed theoretically that by exciting each of three sets of traps in a periodic sequence, nanoparticles could be moved between adjacent traps and could be moved northeast, southwest, clockwise or counterclockwise by drawing down a track of arbitrary length [147]. They experimentally demonstrated controlled particle transport between a pair of C-shaped nanoapertures for polystyrene beads of diameters 200, 390 and 500 nm [148]. They claimed that their plasmonic design opens the door to a new approach for parallel manipulation of nano-objects.

### 3.7 POT with black-Si plasmonic nanostructures

In 2012, a new POT design based on localized plasmonic fields around sharp metallic features was developed [107]. The plasmonic structure consisted of a laser-structured silicon wafer with quasi-ordered microspikes on the surface, coated with a thin silver layer. The authors claimed that the well-formed microspike morphology, coated with silver nanoparticles, showed two orders of magnitude enhancement of the optical trapping force exerted on a 900 nm polystyrene particle, compared to the force obtained with a conventional OT [107]. They mentioned that the roughness of the sharp microspikes was created by silver nanoparticles that were formed on the spikes during the silver deposition [107]. This configuration produced localized plasmonic fields, which enhanced the trapping efficiency. However, since silver is prone to oxidation, to overcome this drawback and simultaneously to minimize the thermal convention in the surrounding medium, Kotsifaki et al. [108, 109] introduced a plasmonic structure based on fs-laser nanostructured silicon samples, coated with thin bilayers of copper/gold. A trapping laser beam at 1070 nm wavelength was employed to trap fluorescent particles of 400 nm diameter above the nanostructured substrates [108]. The authors measured a trapping quality of 0.13 [108], similar to that reported for plasmon-enhanced optical traps [34]. Moreover, they investigated the dependence of the trapping force on the distance from the nanostructured substrates and they verified the exponential distance dependence character of the evanescent plasmon field [108]. A year later [109], the same research group developed an fs POT based on gold-coated black silicon and performed, for the first time, a wavelength-dependent characterization of the trapping process, revealing the resonant character of the trapping quality. They noted a maximum trapping quality value of 0.15 and a resonance wavelength at 975 nm [109]. This design of POT could be a promising new platform for large-scale parallel trapping applications that will broaden the range of precise manipulation of biological specimens.

### 3.8 POT with alternative nanostructure geometries

Additionally, several plasmonic patterns, which are tunable, have been demonstrated to dynamically trap particles, and some of these novel geometries are summarized in this sub-section. Particularly, Tanaka et al. [102, 149] studied theoretically and experimentally the strong trapping potential which can be formed using gold nanoblock pairs. They demonstrated that the size and orientation of the pairs of gold nanoblocks, with 5 nm gaps, determine their surface-plasmon resonance properties and trapping performance in conjunction with the wavelength and the polarization direction of the incident laser [102]. Moreover, they demonstrated optical trapping of 350 nm polystyrene particles with 750 W/cm$^2$ laser intensities [102]. Similarly, 2 years later, they performed two-dimensional mapping of the optical trapping of 100 nm fluorescent polystyrene particles above a gold nanoblock pair that had a 6 nm edge-to-edge separation [149]. They showed that the trap stiffness was enhanced by three orders of magnitude compared with the corresponding one for far-field optical trapping [149]. Furthermore, they found multiple potential wells separated by a distance smaller than the diffraction limit when the incident light polarization was rotated by 90° [149]. This approach indicates the possibility of obtaining super-resolution parallel optical manipulation and opens new channels in designing lab-on-chip devices operated with light.

## 4 Perspectives on the future development of plasmonic optical tweezers

As a rapidly developing research area, the POT has significantly broadened the usefulness and applicability of optical manipulation in the nanoscale regime. Beyond their interest for performing extraordinary experiments, POT may play a key role in new developments and will open up unprecedented opportunities for applications in many scientific fields that will flourish in the coming years. For example, a POT could be integrated with an analytical platform such as a lab-on-a-chip device to provide a promising tool for immobilizing flowing cells, as well as for the detection of biomaterials or viruses in liquid environments – topics that are of interest in several areas of clinical science. Viruses are important biological entities which can affect many aspects of human life. Cheng et al. used two-photon fluorescence techniques with conventional OT to optically manipulate individual HIV-1 viruses [103, 150]. The viruses were labeled with green fluorescent protein, which acts as a fluorescent marker for the virion optical manipulation. However, due to their heterogeneity and their small size, current protocols for their detection often involve a series of steps which are time consuming and require large sample volumes. Integrating an analytical platform with a POT will provide advantages to study the utility of small viruses as reliable predictors by altering their physiology or monitoring the evolution of the disease with high sensitivity and specificity. The potential combination of POT with virology science is a field which has not yet been studied intensively.

Moreover, for detecting and imaging particles which are smaller than the wavelength of light, fluorescent imaging is still the dominant contrast technique. However, the fluorescent technique is limited, because not all molecules or objects are suitable for fluorescent labeling. Additionally, the photobleaching effect will limit the number of photons available for single-molecule detection, thereby also limiting the maximum attainable signal-to-noise ratio. Owing to these obstacles, a possible combination of a POT with high temporal and spatial resolution microscopy techniques is a very attractive prospect. For example, a potential imaging technique for POT will be its combination with interferometry scattering microscopy [151], which pushes the noise performance into the shot noise-limited regime and is not limited by photobleaching. Plasmonic nanostructures [152] with chirality would also enable new applications for POT. Chirality is of major importance in biological systems. For instance, the pharmacological activity of drugs depends mainly on its interaction with biological targets such as proteins, DNA/RNA and bio-membranes. One enantiomer of a chiral drug may be a viable medicine for a particular disease, whereas another enantiomer of the molecule may be not only inactive but could also lead to a detrimental effect [153]. Therefore, the ability for optical manipulation based on chirality will enable innovative applications such as conformational change detection of DNA or proteins and the design or synthesis of new drugs with the possibility of distinguishing their mutant forms or identifying their interactions.

Another potential application of POT is its combination with graphene for studying various systems such as protein secondary structure. Recently, there has been enormous interest in graphene as a platform for several applications, with numerous publications focusing on novel properties offered by this material. In particular, graphene could be used as a biocompatible surface [154] independent of the type of metal used for supporting plasmon resonances. A graphene-gold nanopyramid hybrid OT platform was proposed for trapping and detecting biomolecules [155]. In this study, the graphene acted as a hotspot marker which indicated the intensity and stability of the hotspots [155]. In atomic physics, POT could open new avenues for atom trapping. In particular, the trapping of atoms in a plasmonic optical lattice could enable the study of quantum mechanical interactions between atoms [156]. Finally, a promising next generation of POT, in which their size will be further miniaturized, will make them easier to be used in their own custom setups for on-site investigations, far removed research laboratories.

## 5 Concluding remarks

POT platforms provide versatile nanocontrol of diverse species such as dielectric particles, biomolecules and living cells. In contrast to conventional OT, POT techniques piece together low power and nano-optics principles. We anticipate that POT techniques will not only accelerate progress in scientific research in life sciences, nanoscience and materials sciences, but will also lead to new functional materials, nanomedicines and diagnostic tools. Summarizing, in this work we have reviewed the latest developments in POT based on nanostructures, which have flourished in recent years. We believe that this scientific field will continue to rapidly progress in the future.

**Acknowledgments:** DGK wrote the manuscript under the supervision of SNC. Both authors edited the manuscript. This work was supported by funding from the Okinawa Institute of Science and Technology Graduate University.

**Conflicts of Interest:** The authors declare no conflicts of interest.

# References

[1] Chu S, Bjorkholm JE, Ashkin A, Cable A. Experimental observation of optically trapped atoms. Phys Rev Lett 1986;57:314–7.

[2] Chu S, Hollberg L, Bjorkholm JE, Cable A, Ashkin A. Three-dimensional viscous confinement and cooling of atoms by resonance radiation pressure. Phys Rev Lett 1985;55:48–51.

[3] Bowman RW, Padgett MJ. Optical trapping and binding. Rep Prog Phys 2013;76:026401.

[4] Rodriguez-Sevilla P, Labrador-Paez L, Jaque D, Haro-Gonzalez P. Optical trapping for biosensing: materials and applications. J Mater Chem B 2017;5:9085–101.

[5] Gao D, Ding W, Nieto-Vesperinas M, et al. Optical manipulation from the microscale to the nanoscale: fundamentals, advances and prospects. Light Sci Appl 2017;6:e17039.

[6] Zhang H, Liu KK. Optical tweezers for single cells. J R Soc Interface 2008;5:671–90.

[7] Zhao D, Liu S, Gao Y. Single-molecule manipulation and detection. Acta Biochim Biophys Sin 2018;50:231–7.

[8] Ashkin A. Editorial: light and life. Laser Photonics Rev 2011;5:A7–8.

[9] Ashkin A, Dziedzic JM. Optical trapping and manipulation of viruses and bacteria. Science 1987;235:1517–20.

[10] Ashkin A, Dziedzic JM, Yamane T. Optical trapping and manipulation of single cells using infrared laser beams. Nature 1987;24:769–71.

[11] Ashkin A, Dziedzic JM, Bjorkholm JE, Chu S. Observation of a single-beam gradient force optical trap for dielectric particles. Opt Lett 1986;11:288–90.

[12] Novotny L, Bian RX, Xie XS. Theory of nanometric optical tweezers. Phys Rev Lett 1997;79:645–8.

[13] Ashkin A. Acceleration and trapping of particles by radiation pressure. Phys Rev Lett 1970;24:156–9.

[14] Block SM, Goldstein LSB, Schnapp BJ. Bead movement by single kinesin molecules studied with optical tweezers. Nature 1990;348:348–52.

[15] Yin H, Wang MD, Svoboda K, Landick R, Block SM, Gelles J. Transcription against an applied force. Science 1995;270:1653–7.

[16] Wang MD, Yin H, Landick R, Gelles J, Block SM. Stretching DNA with optical tweezers. Biophys J 1997;72:1335–46.

[17] Kepler J. De cometis libelli tres; Typis Andreæ Apergeri, sumptibus Sebastiani Mylii bibliopolæ Augustani, 1619.

[18] Lebedew P. Untersuchungen über die Druckkräfte des Lichtes. Ann Phys 1901;311:433–58.

[19] Maxwell JC. A treatise on electricity and magnetism, Vol. 2. Oxford, Clarendon Press, 1873.

[20] Marago OM, Jones PH, Gucciardi PG, Volpe G, Ferrari AC. Optical trapping and manipulation of nanostructures. Nat Nanotechnol 2013;8:807–19.

[21] Juan ML, Righini M, Quidant R. Plasmon nano-optical tweezers. Nat Photonics 2011;5:349–56.

[22] Nieto-Vesperinas M, Chaumet PC, Rahmani A. Near-field photonic forces. Phil Trans R Soc Lond A 2004;362:719–37.

[23] Lin S, Crozier KB. Planar silicon microrings as wavelength-multiplexed optical traps for storing and sensing particles. Lab Chip 2011;11:4047–51.

[24] Xu Z, Crozier KB. All-dielectric nanotweezers for trapping and observation of a single quantum dot. Opt Express 2019;27:4034–45.

[25] Xu Z, Song W, Crozier KB. Optical trapping of nanoparticles using all-silicon nanoantennas. ACS Photonics 2018;5:4993–5001.

[26] Mandal S, Serey X, Erickson D. Nanomanipulation using silicon photonic crystal resonators. Nano Lett 2010;10:99–104.

[27] Chen YF, Serey X, Sarkar R, Chen P, Erickson D. Controlled photonic manipulation of proteins and other nanomaterials. Nano Lett 2012;12:1633–7.

[28] MacDonald MP, Spalding GC, Dholakia K. Microfluidic sorting in an optical lattice. Nature 2003;426:421–4.

[29] Kim J, Shin JH. Stable, free-space optical trapping and manipulation of sub-micron particles in an integrated microfluidic chip. Sci Rep 2016;6:33842.

[30] Huang L, Martin OJF. Reversal of the optical force in a plasmonic trap. Opt Lett 2008;33:3001–3.

[31] Al Balushi AA, Kotnala A, Wheaton S, Gelfand RM, Rajashekara Y, Gordon R. Label-free free-solution nanoaperture optical tweezers for single molecule protein studies. Analyst 2015;140:4760–78.

[32] Huang JS, Yang YT. Origin and future of plasmonic optical tweezers. Nanomaterials 2015;5:1048–65.

[33] Shoji T, Tsuboi Y. Plasmonic optical tweezers toward molecular manipulation: tailoring plasmonic nanostructure, light source, and resonant trapping. J Phys Chem Lett 2014;5:2957–67.

[34] Grigorenko AN, Roberts NW, Dickinson MR, Zhang Y. Nanometric optical tweezers based on nanostructured substrates. Nat Photonics 2008;2:365–70.

[35] Righini M, Zelenina AS, Girard C, Quidant R. Parallel and selective trapping in a patterned plasmonic landscape. Nat Phys 2007;3:477–80.

[36] Nichols EF, Hull GF. A preliminary communication on the pressure of heat and light radiation. Phys Rev I 1901;13:307–20.

[37] Nichols EF, Hull GF. The pressure due to radiation. Proc Am Acad Arts 1903;38:559–99.

[38] Ashkin A. Optical trapping and manipulation of neutral particles using lasers. Proc Natl Acad Sci USA 1997;94:4853–60.

[39] Neuman KC, Block SM. Optical trapping. Rev Sci Instrum 2004;75:2787–809.

[40] Harada Y, Asakura T. Radiation forces on a dielectric sphere in the Rayleigh scattering regime. Opt Commun 1996;124:529–41.

[41] Gordon JP. Radiation forces and momenta in dielectric media. Phys Rev A 1973;8:14–21.

[42] Purcell EM, Pennypacker CR. Scattering and absorption of light by nonspherical dielectric grains. Astrophys J 1973;186:705.

[43] Albaladejo S, Marqués MI, Laroche M, Sáenz JJ. Scattering forces from the curl of the spin angular momentum of a light field. Phys Rev Lett 2009;102:113602.

[44] Spesyvtseva SES, Dholakia K. Trapping in a material world. ACS Photonics 2016;3:719–36.

[45] Draine BT. The discrete-dipole approximation and its application to interstellar graphite grains. Astrophys J 1988;333:848–72.

[46] Wright WH, Sonek GJ, Berns MW. Radiation trapping forces on microspheres with optical tweezers. Appl Phys Lett 1993;63:715–7.

[47] Martin OJF, Girard C. Controlling and tuning strong optical field gradients at a local probe microscope tip apex. Appl Phys Lett 1997;70:705–7.

[48] Okamoto K, Kawata S. Radiation force exerted on subwavelength particles near a nanoaperture. Phys Rev Lett 1999;83:4534–7.

[49] Neumeier L, Quidant R, Chang DE. Self-induced back-action optical trapping in nanophotonic systems. New J Phys 2015;17:123008.

[50] Juan ML, Gordon R, Pang Y, Eftekhari F, Quidant R. Self-induced back-action optical trapping of dielectric nanoparticles. Nat Phys 2009;5:915–9.
[51] Hu J, Lin S, Kimerling LC, Crozier K. Optical trapping of dielectric nanoparticles in resonant cavities. Phys Rev A 2010;82:053819.
[52] Descharmes N, Dharanipathy UP, Diao Z, Tonin M, Houndré R. Observation of back action and self-induced trapping in a hollow planar photonic crystal cavity. Phys Rev Lett 2013;110:23601.
[53] Mestres P, Berthelot J, Aćimović SS, Quidant R. Unraveling the optomechanical nature of plasmonic trapping. Light Sci Appl 2016;5:e16092.
[54] Bethe HA. Theory of diffraction by small holes. Phys Rev 1944;66:163–82.
[55] Genet C, Ebbesen TW. Light in tiny holes. Nature 2007;445:39–46.
[56] Wu DY, Zhang M, Zhao LB, Huang YF, Ren B, Tian ZQ. Surface plasmon-enhanced photochemical reactions on noble metal nanostructures. Sci China Chem 2015;58:574–85.
[57] Cottat M, Thioune N, Gabudean AM, et al. Localized surface plasmon resonance (LSPR) biosensor for the protein detection. Plasmonics 2013;8:699–704.
[58] Lertvachirapaiboon C, Baba A, Ekgasit S, Shinbo K, Kato K, Kaneko F. Transmission surface plasmon resonance techniques and their potential biosensor applications. Biosens Bioelectron 2018;99:399–415.
[59] Pang Y, Gordon R. Optical trapping of 12 nm dielectric spheres using double-nanoholes in a gold film. Nano Lett 2011;11:3763–7.
[60] Roxworthy BJ, Bhuiya AM, Vanka SP, Toussaint KC Jr. Understanding and controlling plasmon-induced convection. Nat Commun 2014;5:3173.
[61] Zhang W, Huang L, Santschi C, Martin OJF. Trapping and sensing 10 nm metal nanoparticles using plasmonic dipole antennas. Nano Lett 2010;10:1006–11.
[62] Wang K, Crozier KB. Plasmonic trapping with a gold nanopillar. Chemphyschem 2012;13:2639–48.
[63] Ploschner M, Mazilu M, Krauss TF, Dholakia K. Optical forces near a nanoantenna. J Nanophotonics 2010;4:041570.
[64] Verschueren DV, Pud S, Shi X, De Angelis L, Kuipers L, Dekker C. Label-free optical detection of DNA translocations through plasmonic nanopores. ACS Nano 2019;13:61–70.
[65] Xu Z, Song W, Crozier KB. Direct particle tracking observation and Brownian dynamics simulations of a single nanoparticle optically trapped by a plasmonic nanoaperture. ACS Photonics 2018;5:2850–9.
[66] Jiang Q, Rogez B, Claude JB, Baffou G, Wenger J. Temperature measurement in plasmonic nanoapertures used for optical trapping. ACS Photonics 2019. doi:10.1021/acsphotonics.9b00519. https://arxiv.org/ftp/arxiv/papers/1906/1906.01947.pdf.
[67] Wang K, Schonbrun E, Steinvurzel P, Crozier KB. Trapping and rotating nanoparticles using a plasmonic nano-tweezer with an integrated heat sink. Nat Commun 2011;2:469.
[68] Roxworthy BJ, Toussaint KC Jr. Plasmonic nanotweezers: strong influence of adhesion layer and nanostructure orientation on trapping performance. Opt Express 2012;20:9591–603.
[69] Nicoli F, Verschueren D, Klein M, Dekker C, Jonsson MP. DNA translocations through solid-state plasmonic nanopores. Nano Lett 2014;14:6917–25.
[70] Shi X, Verschueren DV, Dekker C. Active delivery of single DNA molecules into a plasmonic nanopore for label-free optical sensing. Nano Lett 2018;18:8003–10.
[71] Blanchard-Dionne AP, Guyot L, Patskovsky S, Gordon R, Meunier M. Intensity based surface plasmon resonance sensor using a nanohole rectangular array. Opt Express 2011;19:15041–6.
[72] Verschueren D, Shi X, Dekker C. Nano-optical tweezing of single proteins in plasmonic nanopores. Small 2019;3:1800465.
[73] Roxworthy BJ, Ko KD, Kumar A, et al. Application of plasmonic bowtie nanoantenna arrays for optical trapping, stacking, and sorting. Nano Lett 2012;12:796–801.
[74] Ndukaife JC, Kildishev AV, Agwu Nnanna AG, Shalaev VM, Wereley ST, Boltasseva A. Long-range and rapid transport of individual nano-objects by a hybrid electrothermoplasmonic nanotweezer. Nat Nanotechnol 2016;11:53–9.
[75] Kumari P, Dharmadhikari JA, Dharmadhikari AK, Basu H, Sharma S, Mathur D. Optical trapping in an absorbing medium: from optical tweezing to thermal tweezing. Opt Express 2012;20:4645–52.
[76] Kang Z, Chen J, Wu SY, et al. Trapping and assembling of particles and live cells on large-scale random gold nano-island substrates. Sci Rep 2015;5:9978.
[77] Gramotnev DK, Bozhevolnyi SI. Plasmonics beyond the diffraction limit. Nat Photonics 2010;4:83–91.
[78] Liu N, Mesch M, Weiss T, Hentschel M, Giessen H. Infrared perfect absorber and its application as plasmonic sensor. Nano Lett 2010;10:2342–8.
[79] MacDonald KF, Sámson ZL, Stockman MI, Zheludev NI. Ultrafast active plasmonics. Nat Photonics 2008;3:55–8.
[80] Xu T, Wu YK, Luo X, Guo LJ, Plasmonic nanoresonators for high-resolution colour filtering and spectral imaging. Nat Commun 2010;1:59.
[81] Atwater HA, Polman A. Plasmonics for improved photovoltaic devices. Nat Mater 2010;9:205–13.
[82] Ozbay E. Plasmonics: merging photonics and electronics at nanoscale dimensions. Science 2006;311:189–93.
[83] Conway JA, Sahni S, Szkopek T. Plasmonic interconnects versus conventional interconnects: a comparison of latency, crosstalk and energy costs. Opt Express 2007;15:4474–84.
[84] Anker JN, Hall WP, Lyandres O, et al. Biosensing with plasmonic nanosensors. Nat Mater 2008;7:442–53.
[85] Bochenkov VE, Shabatina TI. Chiral plasmonic biosensors. Biosensors 2018;8:120.
[86] Valsecchi C, Brolo AG. Periodic metallic nanostructures as plasmonic chemical sensors. Langmuir 2013;29:5638–49.
[87] Quidant R. Plasmonic tweezers – the strength of surface plasmons. MRS Bulletin 2012;37:739–44.
[88] Volpe G, Quidant R, Badenes G, Petrov D. Surface plasmon radiation forces. Phys Rev Lett 2006;96:238101.
[89] Righini M, Volpe G, Girard C, Petrov D, Quidant R. Surface plasmon optical tweezers: tunable optical manipulation in the femtonewton range. Phys Rev Lett 2008;100:186804.
[90] Sidorov AR, Zhang Y, Grigorenko AN, Dickinson MR. Nanometric laser trapping of microbubbles based on nanostructured substrates. Opt Commun 2007;278:439–44.
[91] Chen KY, Lee AT, Hung CC, Huang JS, Yang YT. Transport and trapping in two-dimensional nanoscale plasmonic optical lattice. Nano Lett 2013;13:4118–22.

[92] Kotnala A, Gordon R. Quantification of high-efficiency trapping of nanoparticles in a double nanohole optical tweezer. Nano Lett 2014;14:853–6.

[93] Pang Y, Gordon R. Optical trapping of a single protein. Nano Lett 2012;12:402–6.

[94] Xu H, Jones S, Choi BC, Gordon R. Characterization of individual magnetic nanoparticles in solution by double nanohole optical tweezers. Nano Lett 2016;16:2639–43.

[95] Han X, Truong VG, Nic Chormaic S. Efficient microparticle trapping with plasmonic annular apertures arrays. Nano Futures 2018;2:035007.

[96] Han X, Truong VG, Thomas PS, Nic Chormaic S. Sequential trapping of single nanoparticles using a gold plasmonic nanohole array. Photon Res 2018;6:981–6.

[97] Saleh AAE, Dionne JA. Toward efficient optical trapping of sub-10-nm particles with coaxial plasmonic apertures. Nano Lett 2012;12:5581–6.

[98] Righini M. Ghenuche P, Cherukulappurath S, Myroshnychenko V, Garcia de Abajo FJ, Quidant R. Nano-optical trapping of Rayleigh particles and *Escherichia coli* bacteria with resonant optical antennas. Nano Lett 2009;9:3387–91.

[99] Tsuboi Y, Shoji T, Kitamura N, et al. Optical trapping of quantum dots based on gap-mode-excitation of localized surface plasmon. J Phys Chem Lett 2010;1:2327–33.

[100] Toshimitsu M, Matsumura Y, Shoji T, et al. Metallic-nanostructure-enhanced optical trapping of flexible polymer chains in aqueous solution as revealed by confocal fluorescence microspectroscopy. J Phys Chem C 2012;116:14610–8.

[101] Shoji T, Saitoh J, Kitamura N, et al. Permanent fixing or reversible trapping and release of DNA micropatterns on a gold nanostructure using continuous-wave or femtosecond-pulsed near-infrared laser light. J Am Chem Soc 2013;135:6643–8.

[102] Tanaka Y, Sasaki K. Optical trapping through the localized surface-plasmon resonance of engineered gold nanoblock pairs. Opt Express 2011;19:17462–8.

[103] Pang Y, Song H, Kim JH, Hou X, Cheng W. Optical trapping of individual human immunodeficiency viruses in culture fluid reveals heterogeneity with single-molecule resolution. Nat Nanotechnol 2014;9:624–30.

[104] Roxworthy BJ, Toussaint KC Jr. Femtosecond-pulsed plasmonic nanotweezers. Sci Rep 2012;2:660.

[105] El Eter A, Hameed NM, Baida FI, et al. Fiber-integrated optical nano-tweezer based on a bowtie-aperture nano-antenna at the apex of a SNOM tip. Opt Express 2014;22:10072–80.

[106] Kang JH, Kim K, Ee HS, et al. Low-power nano-optical vortex trapping via plasmonic diabolo nanoantennas. Nat Commun 2011;2:582.

[107] Kotsifaki DG, Kandyla M, Zergioti I, Makropoulou M, Chatzitheodoridis E, Serafetinides AA. Optical tweezers with enhanced efficiency based on laser-structured substrates. Appl Phys Lett 2012;101:011102.

[108] Kotsifaki DG, Kandyla M, Lagoudakis PG. Near-field enhanced optical tweezers utilizing femtosecond-laser nanostructured substrates. Appl Phys Lett 2015;107:211111.

[109] Kotsifaki DG, Kandyla M, Lagoudakis PG. Plasmon enhanced optical tweezers with gold-coated black silicon. Sci Rep 2016;6:26275.

[110] Garcés-Chávez V, Quidant R, Reece PJ, Badenes G, Torner L, Dholakia K. Extended organization of colloidal microparticles by surface plasmon polariton excitation. Phys Rev B 2006;73:085417.

[111] Min C, Shen Z, Shen FJ, et al. Focused plasmonic trapping of metallic particles. Nat Commun 2013;4:2891.

[112] Lin PT, Chu HY, Lu TW, Lee PT. Trapping particles using waveguide-coupled gold bowtie plasmonic tweezers. Lab Chip 2014;14:4647–52.

[113] Kang Z, Lu H, Chen J, Chen K, Xu F, Ho HP. Plasmonic graded nano-disks as nano-optical conveyor belt. Opt Express 2014;22:19567–72.

[114] Baev A, Furlani EP, Prasad PN, Grigorenko AN, Roberts NW. Laser nanotrapping and manipulation of nanoscale objects using subwavelength apertured plasmonic media. J Appl Phys 2008;103:084316.

[115] Çetin AE, Yanik AA, Yilmaz C, Somu S, Busnaina A, Altug H. Monopole antenna arrays for optical trapping, spectroscopy, and sensing. Appl Phys Lett 2011;98:111110.

[116] Rasmussen MB, Oddershede LB, Siegumfeldt H. Optical tweezers cause physiological damage to *Escherichia coli* and *Listeria* bacteria. Appl Environ Microbiol 2008;74:2441–6.

[117] Shoji T, Mizumoto Y, Ishihara H, et al. Plasmon-based optical trapping of polymer nano-spheres as explored by confocal fluorescence microspectroscopy: a possible mechanism of a resonant excitation effect. Jpn J Appl Phys 2012;51:092001.

[118] Shoji T, Sugo D, Nagasawa F, Murakoshi K, Kitamura N, Tsuboi Y. Highly sensitive detection of organic molecules on the basis of a poly(N-isopropylacrylamide) microassembly formed by plasmonic optical trapping. Anal Chem 2017;89:532–7.

[119] Schuck PJ, Fromm DP, Sundaramurthy A, Kino GS, Moerner WE. Improving the mismatch between light and nanoscale objects with gold bowtie nanoantennas. Phys Rev Lett 2005;94:017402.

[120] Grober RD, Schoelkopf RJ, Prober DE. Optical antenna: towards a unity efficiency near-field optical probe. Appl Phys Lett 1997;70:1354–6.

[121] Crozier KB, Sundaramurthy A, Kino GS, Quate CF. Optical antennas: resonators for local field enhancement. J Appl Phys 2003;94:4632–42.

[122] Fischer H, Martin OJF. Engineering the optical response of plasmonic nanoantennas. Opt. Express 2008;16:9144–54.

[123] Lu Y, Du G, Chen F, Yang Q, Bian H, Yong J, Hou X. Tunable potential well for plasmonic trapping of metallic particles by bowtie nano-apertures. Sci Rep 2016;6:32675.

[124] Jensen RA, Huang IC, Chen O, et al. Optical trapping and two-photon excitation of colloidal quantum dots using bowtie apertures. ACS Photonics 2016;3:423–7.

[125] Berthelot J, Acimovic SS, Juan ML, Kreuzer MP, Renger J, Quidant R. Three-dimensional manipulation with scanning near-field optical nanotweezers. Nat Nanotechnol 2014;9:295–9.

[126] Chen C, Juan ML, Li Y, et al. Enhanced optical trapping and arrangement of nano-objects in a plasmonic nanocavity. Nano Lett 2012;12:125–32.

[127] Ghorbanzadeh M, Jones S, Moravvej-Farshi MK, Gordon R. Improvement of sensing and trapping efficiency of double nanohole apertures via enhancing the wedge plasmon polariton modes with tapered cusps. ACS Photonics 2017;4:1108–13.

[128] Yoo D, Gurunatha KL, Choi HK, et al. Low-power optical trapping of nanoparticles and proteins with resonant coaxial nanoaperture using 10 nm gap. Nano Lett 2018;18:3637–42.

[129] Zehtabi-Oskuie A, Bergeron JG, Gordon R. Flow-dependent double-nanohole optical trapping of 20 nm polystyrene nanospheres. Sci Rep 2012;2:966.

[130] Al Balushi AA, Gordon R. A label-free untethered approach to single-molecule protein binding kinetics. Nano Lett 2014;14:5787–91.
[131] Al Balushi AA, Gordon R. Label-free free-solution single-molecule protein–small molecule interaction observed by double-nanohole plasmonic trapping. ACS Photonics 2014;1:389–93.
[132] Al Balushi AA, Zehtabi-Oskuie A, Gordon R. Observing single protein binding by optical transmission through a double nanohole aperture in a metal film. Biomed Opt Express 2013;4:1504–11.
[133] Gordon R. Biosensing with nanoaperture optical tweezers. Opt Laser Technol 2019;109:328–35.
[134] Hacohen N, Ip CJX, Gordon R. Analysis of egg white protein composition with double nanohole optical tweezers. ACS Omega 2018;3:5266–72.
[135] Kotnala A, De Paoli D, Gordon R. Sensing nanoparticles using a double nanohole optical trap. Lab Chip 2013;13:4142–6.
[136] Kotnala A, Gordon R. Double nanohole optical tweezers visualize protein p53 suppressing unzipping of single DNA-hairpins. Biomed Opt Express 2014;5:1886–94.
[137] Kotnala A, Wheaton S, Gordon R. Playing the notes of DNA with light: extremely high frequency nanomechanical oscillations. Nanoscale 2015;7:2295–300.
[138] Regmi R, Al Balushi AA, Rigneault H, Gordon R, Wenger J. Nanoscale volume confinement and fluorescence enhancement with double nanohole aperture. Sci Rep 2015;5:15852.
[139] Wheaton S, Gordon R. Molecular weight characterization of single globular proteins using optical nanotweezers. Analyst 2015;140:4799–803.
[140] Zehtabi-Oskuie A, Jiang H, Cyr BR, Rennehan DW, Al-Balushi AA, Gordon R. Double nanohole optical trapping: dynamics and protein-antibody co-trapping. Lab Chip 2013;13:2563–8.
[141] Zehtabi-Oskuie A, Zinck AA, Gelfand M, Gordon R. Template stripped double nanohole in a gold film for nano-optical tweezers. Nanotechnology 2014;25:495301.
[142] Aporvari MS, Kheirandish F, Volpe G. Optical trapping and control of a dielectric nanowire by a nanoaperture. Opt Lett 2015;40:4807–10.
[143] Kwak ES, Onuta TD, Amarie D, et al. Optical trapping with integrated near-field apertures. J Phys Chem B 2004;108:13607–12.
[144] Gelfand RM, Wheaton S, Gordon R. Cleaved fiber optic double nanohole optical tweezers for trapping nanoparticles. Opt Lett 2014;39:6415–7.
[145] Sergides M, Truong VG, Nic Chormaic S. Highly tunable plasmonic nanoring arrays for nanoparticle manipulation and detection. Nanotechnology 2016;27:365301.
[146] Zhao Y, Saleh AA, Dionne JA. Enantioselective optical trapping of chiral nanoparticles with plasmonic tweezers. ACS Photonics 2016;3:304–9.
[147] Hansen P, Zheng Y, Ryan J, Hesselink L. Nano-optical conveyor belt, part I: theory. Nano Lett 2014;14:2965–70.
[148] Zheng Y, Ryan J, Hansen P, Cheng YT, Lu TJ, Hesselink L. Nano-optical conveyor belt, part II: demonstration of handoff between near-field optical traps. Nano Lett 2014;14:2971–6.
[149] Tanaka Y, Kaneda S, Sasaki K. Nanostructured potential of optical trapping using a plasmonic nanoblock pair. Nano Lett 2013;13:2146–50.
[150] DeSantis MC, Kim JH, Song H, Klasse PJ, Cheng W. Quantitative correlation between infectivity and Gp120 density on HIV-1 virions revealed by optical trapping virometry. J Biol Chem 2016;291:13088–97.
[151] Ortega Arroyo J, Cole D, Kukura P. Interferometric scattering microscopy and its combination with single-molecule fluorescence imaging. Nat Protoc 2016;11:617–33.
[152] Kong XT, Besteiro LV, Wang Z, Govorov AO. Plasmonic chirality and circular dichroism in bioassembled and nonbiological systems: theoretical background and recent progress. Adv Mater 2018:1801790.
[153] Chhabra N, Aseri ML, Padmanabhan D. A review of drug isomerism and its significance. Int J Appl Basic Med Res 2013;3:16–8.
[154] Pinto AM, Gonçalves IC, Magalhães FD. Graphene-based materials biocompatibility: a review. Colloid Surface B 2013;111:188–202.
[155] Yan Z, Xia M, Zhang P, Xie YH. Self-aligned trapping and detecting molecules using a plasmonic tweezer with an integrated electrostatic cell. Adv Optical Mater 2017;5:1600329.
[156] Chang DE, Thompson JD, Park H, et al. Trapping and manipulation of isolated atoms using nanoscale plasmonic structures. Phys Rev Lett 2009;103:123004.